# Detection of the Missing Baryons in a Warm-Hot Intergalactic Medium

F. Nicastro[1,2], J. Kaastra[3], Y. Krongold[4], S. Borgani[5,6,7], E. Branchini[8], R. Cen[9], M. Dadina[10], C.W. Danforth[11], M. Elvis[2], F. Fiore[1], A. Gupta[12], S. Mathur[13], D. Mayya[14], F. Paerels[15], L. Piro[16], D. Rosa-Gonzalez[14], J. Schaye[17], J.M. Shull[11], J. Torres-Zafra[18], N. Wijers[17], L. Zappacosta[1]

[1] Istituto Nazionale di Astrofisica (INAF) – Osservatorio Astronomico di Roma, via Frascati, Monte Porzio Catone 00078, RM, Italy, email: fabrizio.nicastro@oa-roma.inaf.it
[2] Harvard-Smithsonian Center for Astrophysics, Cambridge, MA, USA
[3] SRON Netherlands Institute for Space Research, Utrecht, The Netherlands
[4] Instituto de Astronomia Universidad Nacional Autonoma de Mexico, D.F., Mexico
[5] University of Trieste Physics Department, Trieste, Italy
[6] INAF – Osservatorio Astronomico di Trieste, Trieste, Italy
[7] Istituto Nazionale di Fisica Nucleare (INFN) – Sezione di Trieste, Trieste, Italy
[8] University of RomaTre Physics Department, Rome, Italy
[9] Princeton University, Department of Astrophysical Science, Princeton, NJ, USA
[10] INAF – Osservatorio di Astrofisica e Scienza dello Spazio di Bologna, Bologna, Italy
[11] University of Colorado, Center for Astrophysics and Space Astronomy (CASA), Boulder, CO, USA
[12] Columbus State Community College, Columbus, OH, USA
[13] Ohio State University, Columbus, OH, USA
[14] Instituto Nacional de Astrofísica, Óptica y Electrónica, Puebla, Mexico
[15] Columbia University, Department of Astronomy, New York, NY, USA
[16] INAF - Istituto di Astrofisica e Planetologia Spaziali. Rome, Italy
[17] Leiden Observatory, Leiden, The Netherlands
[18] Instituto de Astrofísica de La Plata (IALP-UNLP), La Plata, Argentina

**It has been known for decades that the observed number of baryons in the local universe falls about 30-40% short[1,2] of the total number of baryons predicted by Big-Bang Nucleosynthesis (e.g [3]), inferred by density fluctuations of the Cosmic Microwave Background (e.g. [4,5]) and seen during the first 2-3 billion years of the universe in the so called "Lyman-α Forest"[6,7]. While theory provides a reasonable solution to this paradox, by locating the missing baryons in hot and tenuous filamentary gas connecting galaxies, it also sanctions the difficulty of detecting them because their by far largest constituent, hydrogen, is mostly ionized and therefore virtually invisible in ordinary signal-to-noise Far-Ultraviolet (FUV) spectra (e.g. [8,9]). Indeed, despite the large observational efforts, only a few marginal claims of detection have been made so far ([2,10] and references therein). Here we report observations of two highly ionized oxygen (OVII) intervening absorbers in the exceptionally high signal to noise X-ray spectrum of a quasar at redshift >0.4. These absorbers show no variability over a 2-year timescale and have no associated cold absorption, which makes their quasar's intrinsic outflow or host galaxy interstellar medium (ISM) origins implausible. The OVII systems lie instead in regions characterized by large (×4 compared to average[11]) galaxy over-densities, and their number (down to the sensitivity threshold of our data), agrees well with numerical simulation predictions for the long-sought warm-hot intergalactic medium (WHIM). We conclude that the missing baryons in the WHIM have been found.**

Numerical simulations run in the framework of the commonly accepted (Λ-CDM) cosmological paradigm predict that, starting at redshift of $z\sim2$ and during the continuous process of structure formation, diffuse baryons in the intergalactic medium (IGM) condense into a filamentary web (electron densities $n_e \approx 10^{-6}-10^{-4}$ cm$^{-3}$) and undergo shocks that heat them up to temperatures $T\sim10^5-10^7$ K, making their by far largest constituent, hydrogen, mostly ionized (e.g. [8,9]). At the same time, galactic outflows powered by stellar and AGN feedback, enrich these baryons with metals (e.g. [12]). How far from galaxies these metals roam, depends on the energetics of these winds but it is expected that metals and galaxies will be spatially correlated.

This shock-heated, metal-enriched medium, known as WHIM, is made up of three observationally distinct phases: (1) a warm phase, with $T\approx10^5-10^{5.7}$ K, where neutral hydrogen is still present with ion fraction $f_{HI}>10^{-6}$ and the best observable metal ion tracers are OVI (with main transitions in the FUV) and CV (with transitions in the soft X-rays); (2) a hot phase with $T\approx10^{5.7}-10^{6.3}$ K, where $f_{HI}\approx10^{-6}-10^{-7}$ and OVII (with transitions in the soft X-rays) largely dominates metals with ion fractions near unity; and (3) an even hotter phase ($T\approx10^{6.3}-10^7$ K), coinciding with the outskirts of massive virialized groups and clusters of galaxies, where HI and H-like metals are present only in traces (e.g [9]). The warm phase of the WHIM has indeed been detected and studied in detail in the past few years and is estimated to contain an additional $15^{+8}_{-4}\%$ fraction of the baryons (e.g. [1,2] and references therein; Table 1). This brings the total detected fraction to $61^{+14}_{-12}\%$ but still leaves us with a large ($39^{+12}_{-14}\%$) fraction of elusive baryons, which, if theory is correct, should be searched for in the hotter phases of the WHIM. In particular, the diffuse phase at $T\approx10^{5.7}-10^{6.3}$ K should contain the vast majority of the remaining WHIM baryons, and it is traced by OVII. Optimal signposts for this WHIM phase are then OVII He-α absorption lines, which however are predicted to be relatively narrow ($b_{th}(O)\approx20-46$ km s$^{-1}$), extremely shallow (rest-frame equivalent widths $EW \lesssim 10$ mÅ), and rare (Fig. 3). Such lines are unresolved by current X-ray spectrometers and need a signal to noise ratio per resolution element $SNRE \gtrsim 20$ in the continuum to be detected at a single-line statistical significance (i.e. before accounting for redshift trials: see Methods) $\gtrsim 3\sigma$. This requires multi-million second exposures against the brightest possible targets available at sufficiently high redshift ($z \gtrsim 0.3$).

Here we report on the longest XMM-*Newton* Reflection Grating Spectrometer (RGS) observation ever performed on a single continuum target, the X-ray brightest $z>0.4$ blazar 1ES 1553+113 (see Methods). The quality of this RGS spectrum makes it a real goldmine for X-ray absorption-line studies (see Methods and Extended Data – hereinafter ED - Fig. 1 and 2). We detect a number of unresolved absorption lines in both RGSs and at single-line statistical significance ≥2.7–3σ (here, and throughout the paper, we report a range of statistical significance, where the upper boundary is the actual measured single-line statistical significance, while the lower boundary is the measured significance conservatively corrected for observed systematics in the RGS spectrum: see Methods and ED Fig. 3). Particularly, two of these lines are seen at significances of 4.1–4.7σ (Fig. 1a) and 3.7–4.2σ (Fig. 2b) at wavelengths where no Galactic absorption is expected and no instrumental feature is present (ED Fig. 2; see Methods). We identify these lines with intervening OVII He-α absorbers at redshifts $z_1^X$=0.4339±0.0008 (hereinafter System-1) and $z_2^X = 0.3551^{+0.0003}_{-0.0015}$ (hereinafter System-2), respectively (ED Table 1). The statistical significances decrease to 3.5–4σ and 2.9–3.7σ, respectively, after accounting for the number of redshift trials (see Methods). Interestingly, a lower significance (1.7–2σ) absorption line can be modeled at the wavelength where the System-1 OVII He-β is expected (Fig. 1b, and ED Fig. 1 and 2), which increases the "true" statistical significance of System-1 to 3.9–4.5σ (see Methods).

Given the proximity of our two systems with the upper limit $z \lesssim 0.48$ we estimate for the redshift of our target (see Method), we cannot rule out that these two systems be imprinted by material intrinsic to the blazar environment and outflowing from this at speeds of $\lesssim 0.05 - 0.12c$. However a number of reasons make this scenario implausible (see Methods and ED Fig. 4 and Table 2 for details). The identification of System-1 and -2 as genuine WHIM/CGM systems seems much more likely.

By modeling the X-ray data with our hybrid-ionization models (see Methods), we estimate temperatures, oxygen and equivalent H (modulo the metallicity) column densities of System-1 and -2. We obtain: $T_1^X = (6.8^{+9.6}_{-3.6})\times10^5$ K, $N_{O,1}^X = (7.8^{+3.9}_{-2.4})\times10^{15}$ $N_{H,1}^X = (1.6^{+0.8}_{-0.5})(Z/Z_\odot)^{-1}\times10^{19}$ cm$^{-2}$, and $T_2^X = (5.4^{+9.0}_{-1.7})\times10^5$ K, $N_{O,2}^X = (4.4^{+2.4}_{-2.0})\times 10^{15}$, $N_{H,2}^X = (0.9^{+0.5}_{-0.4})(Z/Z_\odot)^{-1}\times10^{19}$ cm$^{-2}$, respectively.

For both systems, these quantities are in good agreement with predictions for the hot phase of the WHIM[8,9,12]. Moreover, the number of detected OVII absorbers is consistent with predictions from a number of models[8,13,14] (Fig. 3).

The identification of WHIM System-1 and -2 is also supported by three independent pieces of evidence (see Methods for details): (1) significant galaxy over-densities at the locations of the two absorbers (ED Fig. 5-7); (2) for System-1, the compatibility (at 2.3σ level) of the HST-COS spectrum of 1ES 1553+113 with the presence of a very broad and shallow (and so physically consistent with OVII), HI Lyman-α absorber at the redshift of the X-ray absorber; (3) for System-2 the presence of two strong (but physically inconsistent with OVII) intervening hydrogen absorbers only 750 km s$^{-1}$ apart and at redshifts consistent with that of the X-ray absorber (Fig. 2,b,c).

Neither of the two strong HI absorbers seen in the HST-COS spectrum at redshifts consistent with that of System-2 (Fig. 2b,), are sufficiently hot to produce the amount of OVII absorption seen in the X-rays (see Methods). This must then be produced by even hotter gas (as indicated by the X-ray fitting), possibly confined between the two colder HI phases, giving rise to undetectable broad HI Lyman-α. For both X-ray absorbers we then use the HST-COS spectrum to derive 3σ upper limits on the column densities of HI and OVI (see Methods).

For System-1 we obtain $N_{HI,1}^{FUV} < 3.9 \times 10^{13}$ cm$^{-2}$ and $N_{OVI,1}^{FUV} < 3.2 \times 10^{13}$ cm$^{-2}$, whereas for System-2: $N_{HI,2}^{FUV} < 3.5 \times 10^{13}$ cm$^{-2}$ and $N_{OVI,2}^{FUV} < 8.1 \times 10^{13}$. Comparing $N_{OVI}^{FUV}$ with the 1σ lower boundary on $N_O^X$, allows us to constrain the minimum needed ionization correction (see Methods) and so to further limit the temperatures of the two systems in the intervals $T_1^X = (0.8 - 1.6) \times 10^6$ K and $T_2^X = (0.5 - 1.4) \times 10^6$ K. Correcting the 3σ upper limits on $N_{HI}^{FUV}$ for the central values of the HI ionization fractions, gives upper limits on the total H column densities of the two systems: $N_{H,1}^{FUV} < 1.4 \times 10^{20}$ cm$^{-2}$ and $N_{H,2}^{FUV} < 9 \times 10^{19}$ cm$^{-2}$, respectively. Finally, comparing these columns with those obtained from the X-ray data, $N_H^X$, we derive 3σ lower limits on the metallicity of the systems: $Z_1^X > 0.1\ Z_\odot$ and $Z_2^X > 0.1\ Z_\odot$ (ED Table 3).

For both systems we assume as upper limit on the average WHIM metallicity (see Methods) the value $Z_{ICM} = 0.2\ Z_\odot$ found in the peripheries (at $r_{500}$) of the intra-cluster medium (e.g. [15]). With metallicity constrained in the $Z \approx (0.1 - 0.2)\ Z_\odot$ interval, we can now use the 68% confidence intervals on the equivalent H column densities to constrain the cosmological mass density of baryons with temperatures in the interval T≈10$^{5.7}$–10$^{6.2}$ K. Parameterizing the lower limit by the inverse of the metallicity in units $0.2Z_\odot$, we obtain $0.002(Z_{0.2})^{-1} < \Omega_b^{10^{5.7} \leq T(K) \leq 10^{6.2}} h^2 < 0.009$, corresponding to $9(Z_{0.2})^{-1}$–40% of the total baryon density measured by Planck[4], in good agreement with predictions[8,9] and potentially sufficient to complete the baryon census (Fig. 4, Table 1).

Finally, theory predicts that metal-enriched WHIM absorbers should lie in the proximity of galaxy over-densities, either in the circum-galactic medium (CGM) of a particular galaxy or in the more diffuse IGM. Consistent with expectations, our photometric redshifts of the r'>23.5 galaxies in the 30x30 arcmin$^2$ field surrounding 1ES 1553+113 indicate that both WHIM System-1 and -2 are found in regions of significant galaxy over-densities (ED Fig. 5-7; see Methods). For System-1 we have a number of spectroscopic redshift confirmations (i.e. within ±900 km s-1 from the absorber: see Methods) showing in particular the presence of a bright (i'=19.6) spiral at only 129 kpc and -15 km s$^{-1}$ from the absorber. For System-2 the only spectroscopically confirmed galaxy, out of only 4 spectroscopic redshifts available, a bright *i'*=20.5 elliptical, lies far (633 kpc and +370 km s$^{-1}$) from the absorber. This may explain the different baryon column density of the two systems: System-1, with its higher column, could be imprinted by the CGM of the nearby spiral, while System-2, with its lower baryon column density, could be produced by a more extended diffuse IGM filament connecting a structure of galaxies.

**Acknowledgements** Based on observations obtained with XMM-Newton, an ESA science mission with instruments and contributions directly funded by ESA Member States and NASA. F.N. and M.E. acknowledge support from NASA grant NNX17AD76G. SM acknowledges NASA grant NNX16AF49G. SB acknowledges financial support from the agreement ASI-INAF n.2017-14-H.0 and the INFN INDARK grant. YK thanks INAOE for the support offered during a sabbatical visit in 2017, and acknowledges support from grant DGAPA-PAPIIT 106518, and from program DGAPA-PASPA. RC acknowledges support from NSF grant AST-1515389.

**Author Contributions** F.N. designed the study (together with co-authors J.K., L.P., S.B., L.Z., S.M., M.E., R.C., F.P. and A.G.), reduced and analyzed the X-ray data, analyzed the FUV data, extracted diagnostics by modeling the X-ray and FUV data, and wrote the paper. Y.K., J.K., L.P., S.M., M.E., F.P., F.F. and A.G. helped with the analysis and modeling of the X-ray spectra. M.D. and F.N. designed and executed the Monte Carlo simulations used to evaluate the statistical significance of the absorbers. J.K. provided the most up-to-date XMM-Newton RGS calibrations and effective area corrections. C.W.D. and J.M.S. reduced the HST-COS data, extracted the G130 and G160 final spectra, and helped with the interpretation of those spectra. Y.K. performed the optical photometric observations and the reduction and analysis of those data, with the help of D.M. and D.R.G. D.R.G. and J.T.Z. provided the galaxies' spectroscopic redshifts. S.B., E.B., J.S., N.W. and R.C. provided hydrodynamical simulations and general theoretical support to the results. In particular J.S. and N.W. provided results from high resolution Eagle simulations and F.P. helped extracting the number density of OVII absorbers from these simulations. All authors contributed equally to the discussion on the results and commented on the manuscript.

**Author Information** Reprints and permissions information is available at www.nature.com/reprints. The authors declare no competing financial interests. Readers are welcome to comment on the online version of the paper. Correspondence and requests for material should be addressed to F.N. (fabrizio.nicastro@oa-roma.inaf.it).

**Figure 1 Intervening absorber at $z_X$=0.4339±0.0008.** Ratios of XMM-*Newton* RGS1+RGS2 data of the blazar 1ES 1553+113 with their local best-fitting continuum model, showing the two OVII He-α (Fig 1a) and He-β (Fig. 1b) absorption lines identifying System-1 at a 'true' statistical significance of 4.2σ. Shaded blue regions indicate the lines detected in the X-rays. Hatched blue intervals in the line histograms represent ±1σ errors (statistical plus 2% systematics). Black curves are best-fitting Gaussians folded through the instrumental RGS line spread function.

**Figure 2 Intervening Absorber at $z_X = 0.3551^{+0.0003}_{-0.0015}$.** Same as Fig. 1, but showing OVII He-α (Fig. 2a) and two HI Lyman-α (Fig. 2b) and –β (Fig. 2c) absorbers only ~750 km s-1 apart and both at redshifts consistent with the X-ray System-2. The X-ray absorber has a "true" statistical significance of 3.1σ. Neither of the two HI absorbers in Fig. 2b,c can be physically associated to the OVII He-α absorber (Fig. 2a), implying the presence of at least three co-located, but physically dramatically different, gaseous phases.

**Figure 3 Agreement between Data and Predictions.** Cumulative number density of OVII He-α intergalactic absorbers per unit redshift with EW greater than a given threshold, versus the rest-frame EW threshold and from a number of different predictions (as labeled)[22,25], compared with the two data points corresponding to our two detections at $z_X$=0.3551 and $z_X$=0.4339. Vertical 1σ error bars are computed using low-number Poisson statistics. The EAGLE results are for the 100 Mpc reference simulation of [14] at *z*=0.1 and an OVII He-α Doppler parameter b=100 km s$^{-1}$ is used in the column density to EW conversion.

**Figure 4 Updated Baryon Census.** Pie diagram of the baryonic components of the local universe. Hatched regions in the external corona indicate the 90% uncertainties on the individual components, and are plotted across one of the two sides of each slice, to show to what extent could each slice be smaller/bigger. The exception is our measurement of the $10^{5.7} \leq T(K) \leq 10^{6.2}$ WHIM component, where the solid area shows the 3σ lower limit on the amount of baryons in these physical conditions, whereas the hatched area fills in the gap and represents our 3σ upper limit (see also Table 1).

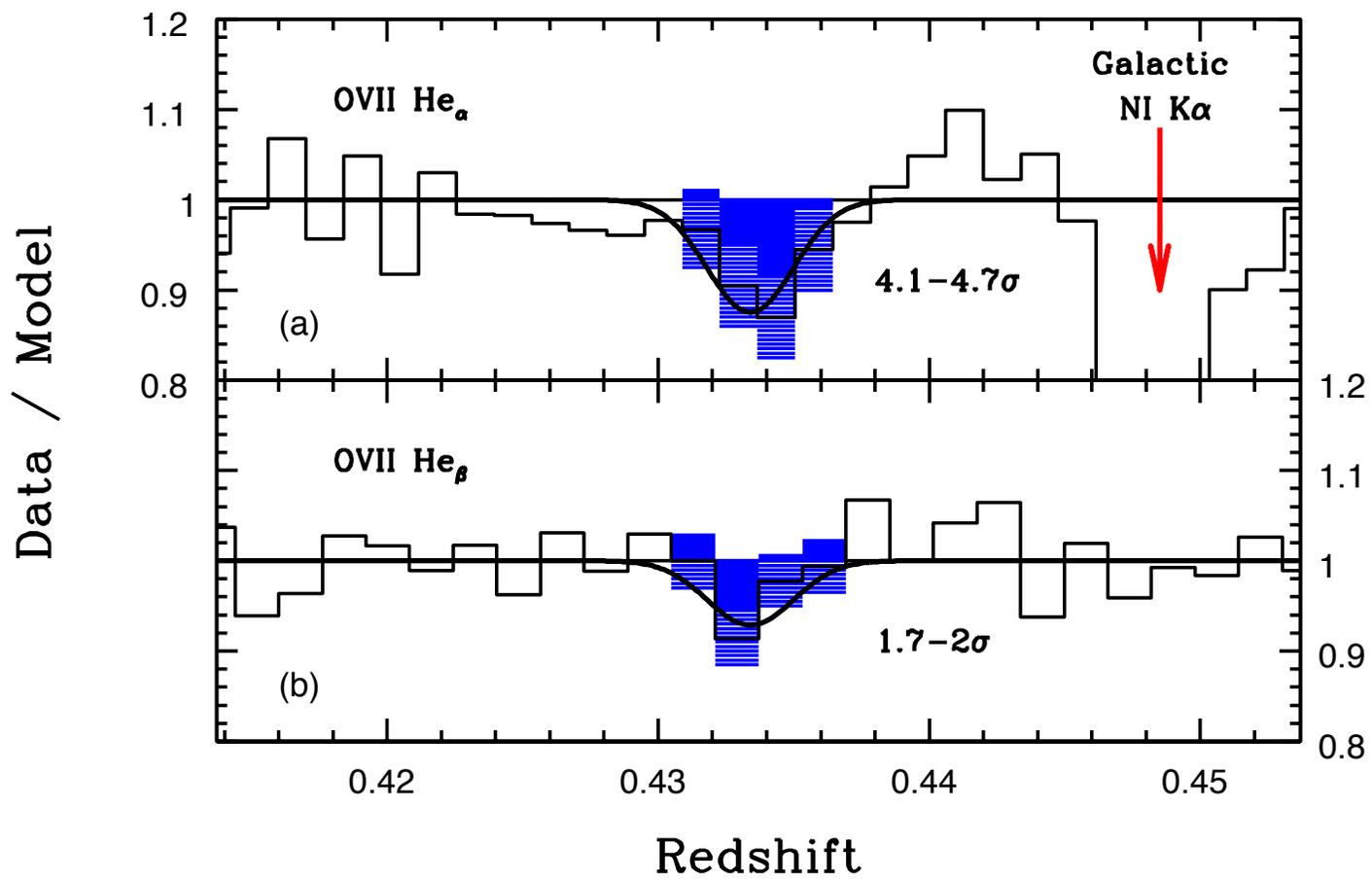

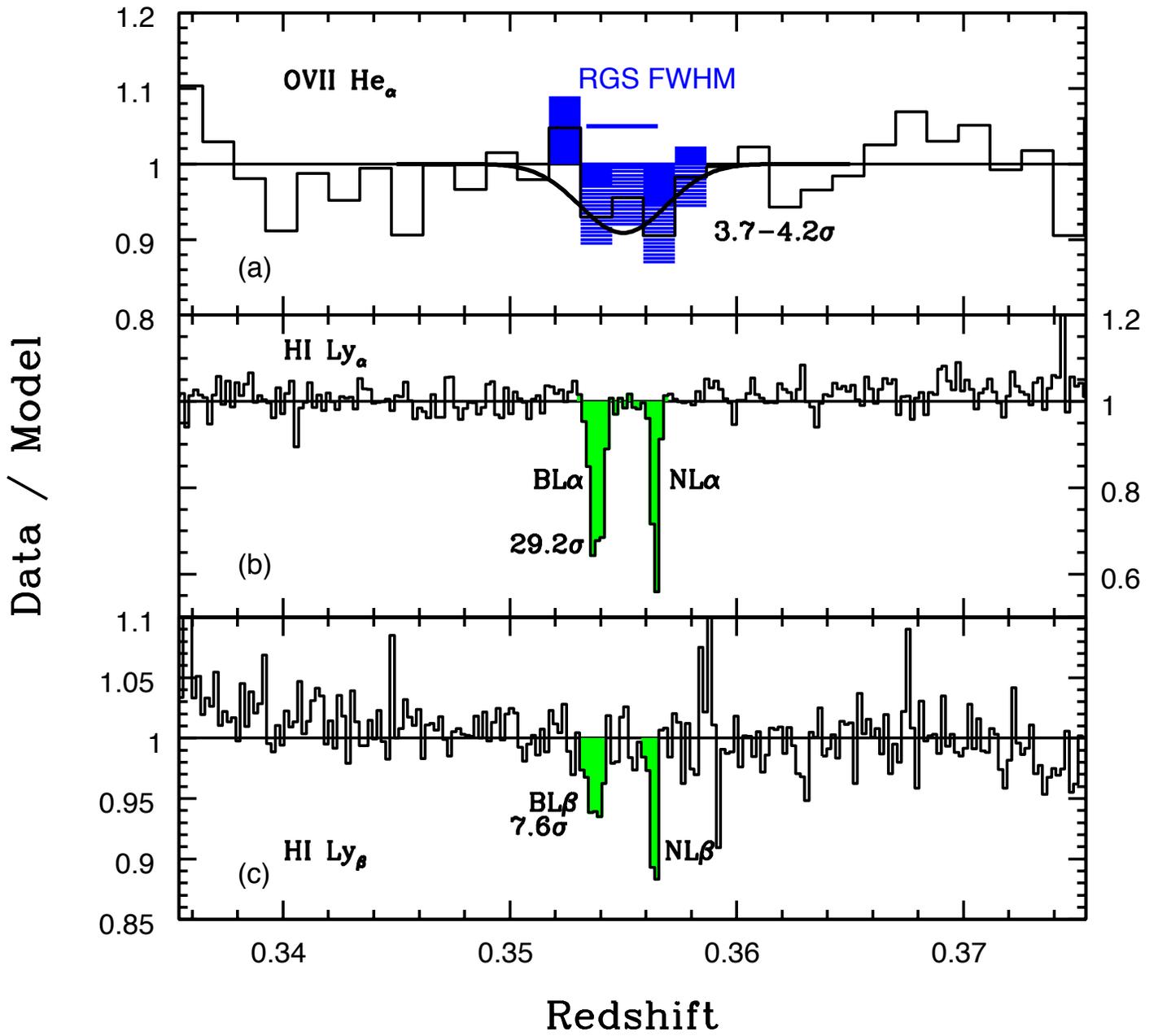

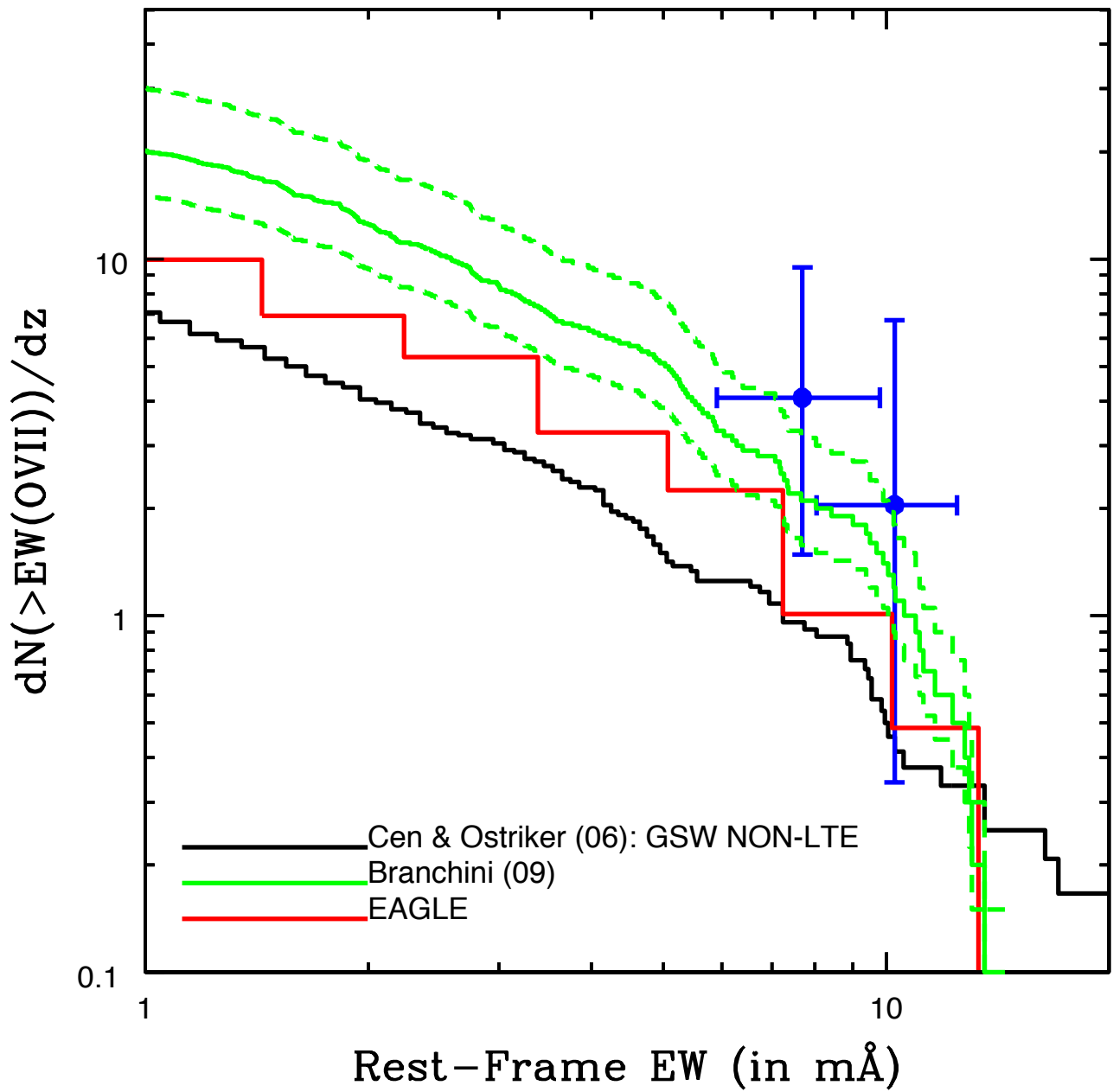

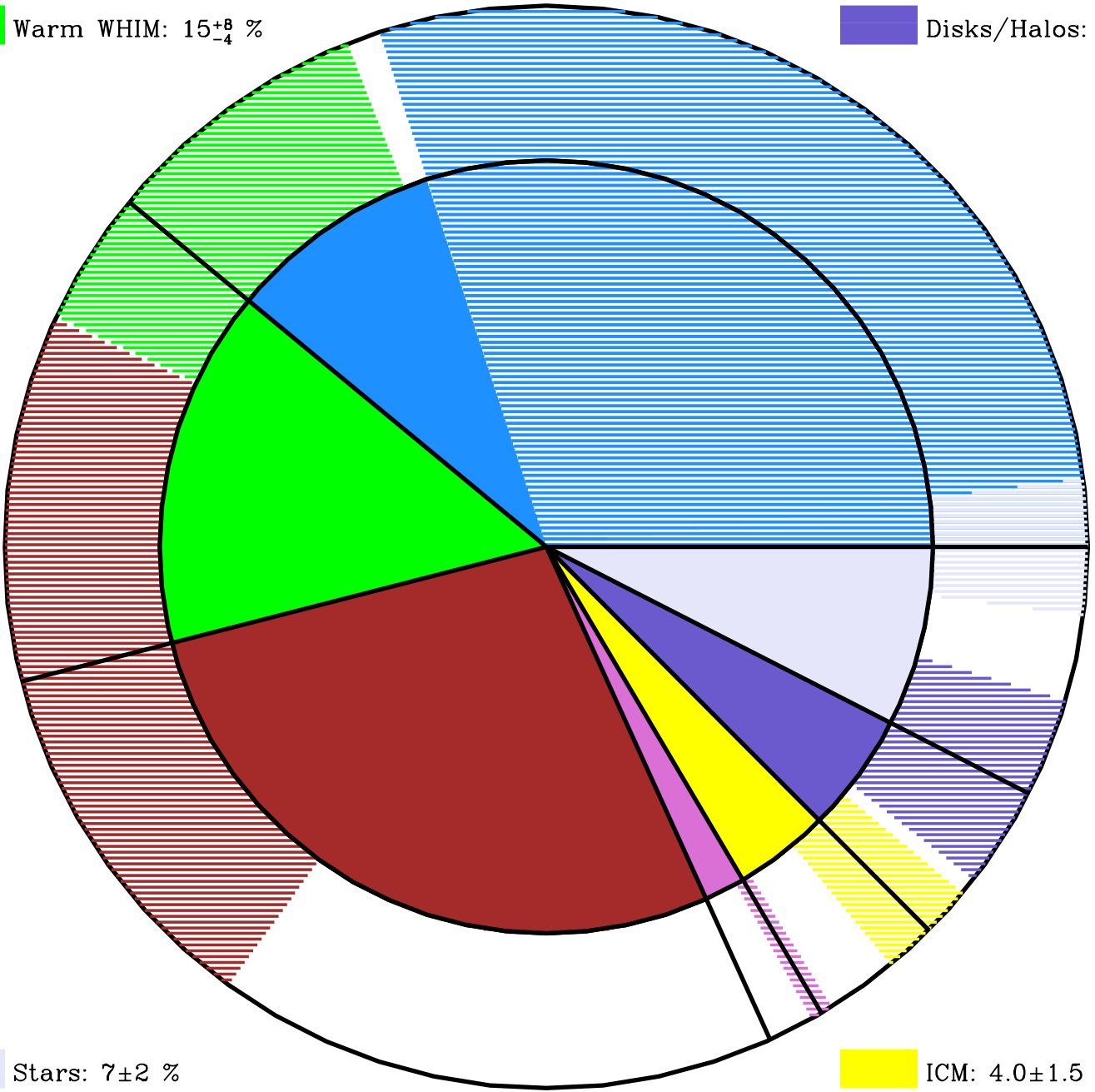

**Table 1 Cosmic Baryon Census at z<0.5.**

| | $\Omega_b h^2$ | $\dfrac{\Omega_b}{\Omega_b^{Plank}}$ |
|---|---|---|
| **Stars in Galaxies** | 0.0015±0.0004 | (7±2)% |
| **Cold Gas in Galaxies** | 0.00037±0.00009 | (1.7±0.4)% |
| **Galaxies' Hot Disks/Halos** | 0.0011±0.0007 | (5±3)% |
| **Hot Intracluster Medium** | 0.00088±0.00033 | (4.0±1.5)% |
| **Photoionized Lyman-α Forest** | 0.0062±0.0024 | (28±11)% |
| **WHIM with $10^5 \leq T(K) < 10^{5.7}$** | $0.0033^{+0.0018}_{-0.0009}$ | $15^{+8}_{-4}$ % |
| **WHIM with $10^{5.7} \leq T(K) \leq 10^{6.2}$** | >0.002 & <0.009 | >9% & <40% |
| **Total** | >0.013 & < 0.026 | >59% & <118% |

## Methods

**Errors.** Throughout the paper uncertainties are quoted at 68% significance, unless explicitly stated.

**Redshift of 1ES 1553+113.** The exact redshift of this blazar is unknown, but a tight spectroscopic lower limit of $z > 0.413$ is imposed by the detection up to this redshift (and close to the long-wavelength end of the HST-COS band-pass) of intervening HI Lyman-α absorption in the HST-COS spectrum of this target (ED Fig. 8; see below). Based on the lack of secure HI Lyman-α detections at $z > 0.413$, the frequency of the HI Lyman-α absorbers (about 1 every 10 Å down to the sensitivity of the highest S/N regions of the HST-COS spectrum) and the decreasing S/N of the HST-COS spectrum at $\lambda > 1750$ Å, we estimate a conservative upper limit of $z \lesssim 0.48$. More recently, a $z = 0.49 \pm 0.04$ redshift estimate has been set based on gamma-ray observations of Extragalactic Background Light absorption[16], consistent with our spectroscopic estimates.

**Statistical Significance.** We define the "single-line statistical significance" of an absorption line as the ratio $EW/\Delta EW$ between the line equivalent width and its negative 1σ error (but see below for a correction due to effective-area systematics in the RGS spectrum). For a blind search of intervening absorbers, this is not the actual statistical significance of the line, unless a prior is used for the absorber redshift. In the absence of such a prior, we estimate the 'true' statistical significance of a given line, by performing Monte Carlo simulations. Each simulation consists in (a) producing 18 RGS1 and RGS2 mock spectra with the same exposures and continuum best-fitting parameters of our 18 XMM observations, (b) co-adding them and (c) fitting the co-added spectra (RGS1 and RGS2) with a continuum model plus an unresolved and negative-only Gaussian, whose position is allowed to vary from the rest-frame position of the transition and its redshifted position at z=0.5 (conservatively assumed to be the blazar redshift, but also, for OVII He-α lines, coinciding with the long wavelength end of the RGSs). We repeat this procedure 10,000 times, and evaluate the chance of detection of a line with single-line-statistical significance greater than a given threshold.

We adopt this 'true' statistical significance for the highest single-line statistical significance X-ray line, and use that line as a prior for other associated X-ray lines, when present. Finally, we evaluate the statistical significance of an X-ray absorption system by adding in quadrature the 'true' statistical significances of the lines of the system.

**XMM-*Newton* RGS spectra of 1ES 1553+113.** XMM-*Newton* observed 1ES 1553+113 for two consecutive cycles and for a total observing time of 1.75 Ms, between July 29, 2015 and September 5, 2015 (for 0.8 Ms; Epoch 1: black spectrum in ED Fig. 4a,b) and between February 1, 2017 and February 22, 2017 (for additional 0.95 Ms; Epoch 2: red spectrum in ED Fig. 4a,b), under a cycle 14 Very Large Program. Four short additional observations performed between September 6, 2001 and July 28, 2014 were also present in the archive, totaling an observing time of 0.096 Ms. We reduced these observations with the latest versions (v. 16.1.0) of the XMM-*Newton* Science Analysis System (SAS) software and Current Calibration Files (CCF) and extracted Reflection Grating

Spectrometer (RGS1 and RG2) spectra and responses by applying the latest calibration corrections. We used the most up-to-date effective area corrections, by switching on the two parameters *withrectification* and *witheffectiveareacorrection* of the SAS tool *rgsrgsrmfge* that generates the RGS response matrices. The second of these parameters, in particular, cures some residual wiggle seen in the spectra of the calibration targets Mkn 421 and PKS 2155-304 at wavelength ~29–32 Å (ED Fig. 2). We also added a 2% systematic uncertainty to the RGS counts, as estimated by the team of RGS calibrators lead by co-author J.K. The final, cleaned, RGS spectra have an exposure time of 1.85 Ms, and are shown in ED Fig. 1.

The RGS spectrum of 1ES 1553+113 has SNRE=33 at $\lambda$=23.5–30.2 Å, and *SNRE*=23 at $\lambda$=21.6–23.5 Å (where only one of the two RGSs is present) and $\lambda$=30.2–32 Å (the long-wavelength end of the RGS band pass). This implies a 90% sensitivity of *EW*≥3.5 mÅ and *EW*≥4.5 mÅ to intervening OVII He-α lines in the two redshift intervals 0.08≤$z$≤0.4 and $z \in$ (0,0.08) ∪ (0.4,0.5), respectively. The 8–33 Å RGS spectrum shows a number of narrow (unresolved) line-like negative features (ED Fig. 1), eight of which are identifiable with Galactic absorption (marked and labeled in blue in ED Fig. 1), namely (using half RGS *FWHM* resolution element as errors on positions): NeX Lyman-α ($\lambda$=12.131±0.035 Å), NeIX He-α ($\lambda$=13.460±0.035 Å), OVII He-α ($\lambda$=21.586±0.035 Å), OIV Kα + OI Kγ ($\lambda$=22.701±0.035 Å), OII Kα ($\lambda$=23.339±0.035 Å), OI Kα ($\lambda$=23.512±0.035 Å), NVI Kα ($\lambda$=28.772±0.035 Å) and NI Kα ($\lambda$=31.281±0.035 Å). Two additional unresolved absorption lines are detected in both RGSs at combined single-line statistical significances of 4.1–4.7σ (Fig. 1a and ED Fig. 1, 2 and Table 1) and 3.7–4.2σ (Fig. 2a and ED Fig. 1, 2 and Table 1), at wavelengths where (1) no Galactic absorption is expected and (2) neither of the two spectrometers is affected by instrumental features due to cool-pixels in the dispersing detector (ED Fig. 2). These are the lines here identified as intervening WHIM OVII He-α at $z_1^X$=0.4339±0.0008 (System-1) and $z_2^X = 0.3551^{+0.0003}_{-0.0015}$ (System-2; ED Table 1 and Fig. 1, 2). An additional lower significance (1.7–2σ) line is detected at a $\lambda$=26.69±0.09 Å, and is identifiable as OVII He-β at a redshift consistent with $z_1^X$ (Fig. 1b, ED Fig. 1, 2 and Table 1).

**Additional Systematics in the RGS spectrum of 1ES 1553+113.** In addition to the 2% systematics estimated by the team of RGS calibrators for any RGS spectrum, we look for the presence of any systematics specific to our co-added 1.85 Ms RGS spectra of 1ES 1553+113, by performing the following Monte Carlo test. We re-fit the RGS spectra of 1ES 1553+113 by adding to the best-fitting continuum plus absorption line models shown in ED Fig.1, an unresolved Gaussian (whose normalization is allowed to be either positive or negative) at a random position in the $\lambda$=8–33 Å range, and evaluate the single-line statistical significance of the line (assumed to be negative for emission lines and positive for absorption lines). There are about 400 RGS resolution elements in this wavelength range, and we therefore repeat this operation 1,000 times. ED Fig. 3 shows the distribution of measured single-line significances (black histogram) and compares it with the expected distribution for a Normal distribution with standard deviation of unity (red curve). The data distribution is symmetric (indicating that any systematics is also acting symmetrically) but slightly flatter than the red curve in ED Fig. 3. We think this is due to uncertainties in the RGS effective areas at the wavelengths corresponding to cool pixels in the dispersing detectors. This effect can indeed be seen in ED Fig. 2, where the residual excesses or deficits of counts in the data at, e.g., $\lambda$~27.7 Å, 30.2–3.3 Å and 31.1–31.2 Å, all correspond to strong effective area instrumental feature (red and black curves). The Normal distribution that best fits our Monte Carlo results, has a standard deviation of 1.15 (green curve in ED Fig. 3), which should then be used to correct the statistical significance of lines detected at wavelengths affected by cool-pixels. The absorption lines reported here are all found in relatively clean effective area spectral regions. However, to be conservative, for each line we use a range of statistical significances, whose lower and upper boundaries are the measured single-line statistical significances corrected for effective-area-induced systematics and the actual single-line significance, respectively. Such single-line statistical significance intervals are then propagated to "true" statistical significance intervals, by allowing for redshift trials as detailed above.

**X-Ray Diagnostics: Temperature and Baryon Column Density.** We use our hybrid-ionization models (i.e. models of collisionally-ionized gas perturbed by photoionization, at a given redshift, by the meta-galactic radiation field)[17], to characterize the temperature $T^X$ and the equivalent H column density $N_H^X$ (modulo the absorber absolute metallicity) of the X-ray absorbers. In our models we use relative metallicities from [18]. Our spectral WHIM models include more than 3,000 line transitions in the soft X-ray band (E≈0.1–2 keV). For each transition, line optical depth and Voigt profile are self-consistently computed for couples of values of the gas temperature (i.e. ionization structure) and equivalent H column density, which are let free to vary in the fit. At typical WHIM conditions, the temperature of these models are well constrained by the detected ion transitions, as well as the upper limits that can be set on the presence of lower- or higher-ionization ion transitions. For example, in the 8–33 Å band covered by the RGSs and at the temperature where OVII is the dominant ion of oxygen, the low-$T$ boundary is mostly set by the non-detection of K-shell transitions of OIV-VI and M-shell transitions of FeVII-XVI, while the high-$T$ boundary is set by the absence of the K-shell transitions of OVIII and NeIX-X and the L-shell transitions of FeXVII-XXIV.

We note that our temperature estimates depend on our assumption of gas in collisional ionization equilibrium, perturbed by the meta-galactic ionizing radiation field. This assumption may not be valid in particularly low-density regions, where the post-shock electron-ion relaxation timescale is longer than the Hubble time (e.g. [19]), and could yield to an over-estimate of the actual electron temperature[19]. However, even for a factor of 2 difference in electron temperature, the fraction of our tracer ion OVII at its peak temperature interval $T \approx 5 \times 10^5 – 2 \times 10^6$ K, is not significantly affected by this effect (see e.g. Fig. 4 of [19]), which make our estimates of oxygen columns densities virtually independent on the exact ionization state of the gas.

The *EW*s of the detected OVII lines, together with the limits on the *EW*s of higher- and lower-ionization oxygen lines, constrain the total oxygen column densities of the X-ray absorbers, and so, modulo their metallicity (which we assume to be solar in the fit), their equivalent H column densities.

Finally, we use our hybrid-ionization models[17], to derive ionization fractions of HI, OVI and OVII in a given temperature interval and for a gas volume density of $n_e=10^{-5}$ cm$^{-3}$ (at densities $n_e>10^{-4}$ cm$^{-3}$, photoionization plays a negligible role, and the gas is virtually in collisional-ionization equilibrium, whereas for $n_e \leq 10^{-4}$ cm$^{-3}$ the uncertainties in the ionization fractions of HI, OVI and OVII are dominated by the uncertainties on the temperature of the absorber rather than its volume density).

**HST-COS spectrum of 1ES 1553+113.** 1ES 1553+113 was observed with the HST Cosmic Origin Spectrograph (COS) a first time on September 22$^{nd}$, 2009 for 3.1 ks and 3.8 ks, with the G130M ($\lambda=1135–1480$ Å) and G160M ($\lambda=1400–1795$ Å) gratings, respectively. This spectrum has a signal to noise per $\Delta\lambda \approx 0.08$ Å resolution element *SNRE*~23 and was published by [20], who reported the detection of 42 intervening IGM systems at $z \lesssim 0.4$. A second spectrum was collected almost two years later, on July 24$^{th}$, 2011, with exposures of 6.4 ks (G130) and 8 ks (G160), which almost doubled the *SNRE* of the 2009 spectrum over its entire G130M+G160M bandpass.

Here we present a targeted analysis of the full HST-COS spectrum of 1ES 1553+113, aimed to specifically search for HI and/or OVI counterparts to the two intervening OVII absorbers identified in the XMM-*Newton* RGS spectra. We do this by using the fitting package *Sherpa* of the Chandra Interactive Analysis of Observation (CIAO) software, to model the normalized HST-COS spectrum with negative Gaussian functions. When lines are seen in the HST-COS spectrum, we leave all the Gaussian parameters (namely, position, width and normalization - or *EW*) free to vary in the fit and let the fitting-routine find the best-fitting parameters by minimizing the statistics with the method *moncar* in *Sherpa*. When more than one transition from the same ion is present and the association is secure, we model the absorption lines by linking their relative positions and widths to the relative rest-frame positions. When lines are not seen, but an EW upper limit is needed for a line with position and width bounded by physically motivated limits (i.e. the redshifts of the two intervening X-ray systems for the line positions, or their temperatures for the line widths), we set these limits as the minimum and maximum values of the Gaussian position and width during the fit, and let the fitting routine find the best-fitting Gaussian normalization within such limits. We then compute the 3σ upper limit on the normalization of such best-fitting Gaussian by using the *Sherpa* task *conf*.

When resolved lines of HI are available, we use the empirical correction of [21] to derive the temperature of the FUV gas: $b_{HI}^{th} = \sqrt{\frac{2kT}{m_p}}$, and $b_{HI}^{th} \approx b_{HI}^{obs}/1.2$. Analogously, we use the inverse relationship to set boundary conditions to the width of an HI (or OVI, factoring the O/H mass factor in) line, when evaluating the 3σ upper limit on the *EW* of such a line.

Following [22], we derive ion column densities of FUV lines by running curve-of-growth analyses of the measured line *EW*s. When more than two transitions from the same ion are present, we check for common solutions in the ion column density versus Doppler parameter plane (e.g. Lyman-α and –β lines near System-2: see below), and then use the estimated temperature to further constrain the ion column density.

**HI and OVI Counterparts of X-ray System-1 and -2.** The HI Lyman-α counterpart to the X-ray System-1 falls just longwards of the strong NiII($\lambda$1741.55) line from our Galaxy's interstellar medium (ISM). This NiII line was misidentified in [20] as an HI Lyman-α absorber at *z*=0.43261. In our re-analysis of the HST-COS spectrum of 1ES 1553+113, we simultaneously fit the 6 available NiII lines ($\lambda$=1317.22 Å, $\lambda$=1370.13 Å, $\lambda$=1454.84 Å, $\lambda$=1709.40 Å, $\lambda$=1741.55 Å and $\lambda$=1751.9 Å). We then add a negative Gaussian to the model, to search for an HI Lyman-α absorber at redshifts within ±1σ from $z_1^X$ and thermal width bounded by the ±1σ uncertainty on $T_1^X$. The fitting routine finds a weak (2.3σ) and broad ($b_{HI}^{obs} \approx 220$ km s$^{-1}$) line (ED Table 1), which we use to set an upper limit on the EW of the HI Lyman-α associated with System-1. Analogously, we use the boundary position and width conditions imposed by the redshift and temperature of System-1, to set an upper limit to the EW of the OVI($\lambda$1031.93) line of System-1 (ED Table 1).

In System-2, two HI Lyman-α (Fig. 2, 2$^{nd}$ panel) and Lyman-β (Fig. 2, 3$^{rd}$ panel) absorbers are present at redshifts consistent with that of the OVII He-α absorber (Fig. 2, top panel): $z_{FUV}$ = 0.35383±0.00001 and $z_{FUV}$=0.35642±0.00001. However, the HI lines at $z_{FUV}$=0.35642 are too narrow ($b_{th}$(H)≈32 km s$^{-1}$, implying *T*≈6x10$^4$ K[21]) to be even tentatively associated to the OVII-bearing gas. For the broader HI absorber (hereinafter BHI) at $z_{FUV}$ = 0.35383±0.00001, we estimate $T_2^{BHI} = (3.5_{-0.7}^{+0.8}) \times 10^5$ K[21], only marginally (2σ) consistent with the X-ray estimate. By combining the

Doppler widths with the observed line *EW*s, we obtain $N^{BHI}_{HI,2} = (3.7 \pm 0.1) \times 10^{14}$ cm$^{-2}$, $N^{BHI}_{OVII,2} = (5.2^{+3.5}_{-2.0}) \times 10^{15}$ cm$^{-2}$ and a 3σ upper limit $N^{BHI}_{OVI,2} < 3.5 \times 10^{13}$ cm$^{-2}$. Factoring the ionization fractions in, we obtain a solid total H column density for BHI: $N^{BHI}_{H,2} = (3.2 \pm 1.0) \times 10^{20}$ cm$^{-2}$. However, the O columns derivable from OVII and OVI, are >3σ inconsistent with each others: $N^{BHI}_{O,2}(OVII) = (7.7^{+5.7}_{-3.8}) \times 10^{15}$ cm$^{-2}$ and $N^{FUV}_{O,2}(OVI) < 0.7 \times 10^{15}$ cm$^{-2}$. This, together with the X-ray-FUV temperature inconsistency, imply that (in the collisional ionization equilibrium hypothesis that we adopt here) neither of the two strong HI absorbers at redshifts consistent with $z^X_2$, can be physically associated to System-2. Following the same procedure as for System-1, we therefore set 3σ upper limits to the *EW*s of the HI Lyman-α and OVI(λ1031.93) transitions (ED Table 1).

**FUV-X-ray diagnostics: refined Temperature and Metallicity.** We use the upper limits on the *EW*s of the OVI(λ103193) transition in System-1 and -2, to refine the allowed OVI ionization fraction intervals and so the lower boundaries of the allowed temperature intervals. Namely, we estimate an upper limit on the OVI ionization fraction by dividing the FUV-derived OVI column density upper limit, by the 1σ lower boundary on the X-ray-derived oxygen column density: $f_{OVI} = N_{OVI}/N_O < N^{FUV}_{OVI}/[N^X_O - (\Delta N^X_O)_-]$. For System-1 this gives revised temperature $T^X_1 = (0.8 - 1.6) \times 10^6$, and ionization fractions $f^{HI}_1 = (2.3 \pm 1.0) \times 10^{-7}$, $f^{OVI}_1 = (0.0041 \pm 0.0018)$ and $f^{OVII}_1 = (0.811 \pm 0.098)$. For System-2 we obtain a revised temperature $T^X_2 = (0.5 - 1.4) \times 10^6$, and ionization fractions $f^{HI}_1 = (4.0 \pm 2.5) \times 10^{-7}$, $f^{OVI}_1 = (0.019 \pm 0.016)$ and $f^{OVII}_1 = (0.855 \pm 0.060)$.

We then use these refined ionization corrections and the upper limits on the *EW*s of the HI Lyman-α transition, to derive 3σ upper limits to the FUV equivalent H column densities, and so 3σ lower limits on the metallicity of the systems. We do this by first dividing the 3σ upper limit on the HI column density, by the central value of the revised HI ionization fraction, $N^{FUV}_H = N^{FUV}_{HI}/f_{HI}$, and then comparing this with the central value of the X-ray-estimated equivalent H column: $Z/Z_\odot = N^X_H/N^{FUV}_H$. As an upper limit on the mean metallicity of the systems, we instead assume the value $Z_{ICM} = 0.2\, Z_\odot$ found in the peripheries (at $r_{500}$) of the intra-cluster medium (e.g. [15]). However, it is possible that the metal distribution in the IGM is inhomogeneous, and in particular it has been argued that oxygen absorbers can arise from relatively over-enriched ($Z \sim 0.5\, Z_\odot$) regions (e.g. [12,13]). Should this be the case for one or both of our OVII absorbers, this would affect (by linearly lowering it) our lower limit on the cosmological mass density of baryons in the WHIM (see below).

**Cosmological Mass Density.** Following Schaye[23], we estimate the cosmological mass density of baryons in the $10^{5.7} \le T(K) \le 10^{6.2}$ WHIM, by using the formula $\Omega_b h^2 (10^{5.7} \le$

$T(K) \leq 10^{6.2}) = \left(\frac{1}{\varrho_c}\right)\left(\frac{m_p \Sigma_i N_H^i}{(1-Y)d}\right)$, where (1–Y) is the hydrogen mass fraction (whose inverse is taken to be ~1.3), $\varrho_c$ is the universe critical density, $N_H^i$ (for $i$=1,2) are the estimated equivalent H column densities for System-1 and 2, and *d* is the available path-length, which, after factoring the reduction of available RGS band-pass due to both Galactic absorption lines and detector cool-pixels, is $\Delta z$=0.42, or *d*=1528 Mpc comoving.

**Optical Data of 1ES 1553+113.** We observed the field of 1ES 1553+113 with the OSIRIS camera at the 10m Gran Telescope Canarias (GTC). We performed a 4x4 mosaic observation centered on 1ES 1553+113, to cover a 30x30 arcmin² field, in the 5 Sloan Digital Sky Survey (SDSS) bands, *u'*, *g'*, *r'*, *i'* and *z'*. Our survey is flux-limited to *r'*~23.5 (factor of 4 deeper than the available SDSS data), corresponding to an absolute r' magnitude of -18.9 at *z*=0.5. We reduced the data using IRAF, along with the *gtcmos* package (http://www.inaoep.mx/~ydm/gtcmos/gtcmos.html). We detect galaxies and perform photometry using *SExtractor*. Following Brescia and collaborators ([24]), we then derive photometric redshifts using *photoraptor* with a combination of the four band colors plus the pivot magnitude in the *r'* band. Our photometric redshifts have an accuracy of $\Delta z \approx 0.07$ in the interval $z \approx 0.15$–0.6.

The histogram in ED Fig. 5, is built by considering photo-z-identified galaxies in cylindrical volumes with base radius 0.5 Mpc and line of sight depth $\Delta z$=0.07, at each redshift bin. Galaxies circled in yellow and cyan in ED Fig. 6 and 7, respectively, are photo-z identified galaxies in cylindrical volumes with base radii 0.5 and 1.75 Mpc and line-of-sight depth $\Delta z$=0.07, centered at the redshifts of System-1 and -2. According to our photometric redshifts, both system-1 and -2 seat at the center of a large concentration of galaxies: (12,54) and (8,72) galaxies with *r'*>23.5 are found in cylindrical volumes with base radii (0.5,1.5) Mpc and line-of-sight depth $\Delta z$=0.07, in the redshift bins of System-1 [where only ~(3,27) are expected[11]] and System-2 [only ~(2,18) expected[11]], respectively.

We also have a number of spectroscopic redshifts, taken with OSIRIS-MOS at GTC and with the GMOS at the Gemini North Telescope. Data have been reduced and analyzed with IRAF. Redshifts are present for 44 galaxies in the 5.5x5.5 arcmin² field around 1ES 1553+113 (35 new measurements and 6 from [25]). For System-1, 13 of our photo-z-identifed galaxies have spectroscopic redshift and for 8 of them the identification is confirmed (i.e. redshifts within ±900 km s⁻¹ from the absorber: solid thick circles in ED Fig. 6). For System-2 only 4 of our photo-z-identified galaxies have spectroscopic redshifts and only one of these four identifications is confirmed (solid thick circle in ED Fig. 7), at a velocity of +370 km/s from the absorber's redshift.

**On the Implausibility of Intrinsic Absorption.** Given the proximity of our two systems with the upper limit $z \lesssim 0.48$ we estimate for the redshift of our target, we cannot rule out that these absorbers be imprinted by material intrinsic to the blazar environment and outflowing from this at speeds of $\lesssim 0.05 - 0.12c$. However a number of reasons (also discussed for an analogous case by [17], section 7.6.1), make this scenario implausible. Here we review some of these reasons for the specific cases of System-1 and -2.

While photo-ionized outflows are commonly seen in type-1 Seyferts and quasars (the so called "warm absorbers" – WAs –, e.g. [26] and references therein), these systems are always seen in multiple species (both in the FUV and X-rays), due to the smoother ion fraction distribution in photo-ionized versus collisionally ionized gas (compare, e.g., top and bottom panels in Fig. 1 and 2 of [27]). Our System-1 and -2, instead, are only seen in OVII, suggesting that electron-ion collisions are the main mechanism of ionization of this gas. Moreover, WAs have typical outflow velocities of only few hundreds to few thousands km s$^{-1}$, whereas even for our System-1 the implied outflow velocity could be as high as 15000 km s$^{-1}$. Finally, WAs have equivalent hydrogen column densities in the range $N_H \sim 10^{20} - 10^{23}$ cm$^{-2}$, at least an order of magnitude higher than observed here[26]. Indeed, no such system has ever been confirmed in blazars, and is exactly the intrinsic featurelessness of their spectra (together with their relatively high X-ray fluxes), which makes blazars particularly suited for IGM-absorption X-ray experiments (e.g. [2] and references therein). Early reports of X-ray absorbers in blazars[28,29] have not been confirmed by later, higher spectral resolution observations (e.g., [30-32]). The only absorption lines detected in high resolution, high S/N X-ray spectra of about a dozen of blazars are either from our own Galaxy's disk and/or halo (e.g. [22,33,34]) or claims of intergalactic WHIM ([2] and references therein). The only two exceptions reported are a misidentification of an OVII He-α line at a redshift consistent with that of Mkn 421[17,35], which is instead a Kβ transition of OII from our own Galaxy[22], and a transient OVIII Kα absorber at the redshift of the blazar H 2356-309[36].

The OVIII Kα absorber reported by Fang and collaborators[36] has an OVIII column two orders of magnitude larger than those reported here for our two OVII absorbers and, perhaps more importantly, it has a transient nature, appearing in only one out of the five consecutive ~80 ks *Chandra* observations of H 2356-309 performed in September 2008. This is not surprising: any gaseous material intrinsic to the AGN environment must experience strong photo-ionization by the quasar's radiation field, which in X-rays varies on timescales as short as few hundred seconds. Consistently the ionization degree of warm absorbers is often seen to vary on timescales from few ks (e.g. [37]) to months (e.g. [38]). On the contrary, no change is seen in our absorbers over the two years elapsed between the first and the second half of our observing campaign (ED. Fig. 4 and Table 2, where we also report the EWs of the strong Galactic line of NI during the two epochs, for comparison), despite the orders of magnitude change commonly experienced by the beamed X-ray luminosity of our target (e.g. [39]). The lack of variability of our OVII absorbers would require electron volume densities of these systems $n_e \lesssim 10^2$ cm$^{-3}$, at least 2 orders of magnitude lower than the lowest limit estimated for warm absorbers (e.g. [38]). An intrinsic AGN outflow origin for our System-1 (and -2), seem then unlikely.

Such low densities are consistent with the disk or moderately extended halo of the blazar's host galaxy, however, not only in such a case only our System-1 could be associated to the blazar's host galaxy (whose redshift would then coincide with that of System-1), but strong OI and OII Kα absorbers should also be seen (as in our own Galaxy, e.g. [22] and ED Fig. 1), and are not.

We conclude that a much more plausible explanation for our System-1 and -2, is intervening absorption by diffuse WHIM or the CGM of intervening galaxies.

**Ruling out Absorption by a thick-disk of an intervening galaxy.** Our OVII System-1 and -2, could in principle be due to absorption by highly ionized material in the thick-disk or halo of an intervening galaxy with impact parameter < 100 kpc (for the brightest spirals). In this case, the derived equivalent H column densities should not be used as described above, to derive a cosmological mass density of hot baryons in the universe. However, there is no $r'>23.5$ galaxy with confirmed impact parameter < 100 kpc from our two absorbers. Moreover, even if small galaxies, fainter than our survey's limit, were present at such small impact parameters, their thick disks and/or halo would also contain a large amount of cool matter with high fractions of neutral and mildly ionized metals. Strong OI and OII absorption is seen, for example, at high galactic latitude in our own Galaxy[22], as well as around the disk (galaxy-absorber impact parameters ~6-100 kpc) of external galaxies[40]. Nicastro and collaborators[22], by comparing lines of sight through the disk of our Galaxy and high galactic latitude lines of sight, show that a cool (T~3000 K) ionized medium traced by OI and OII absorbers (also seen here in the spectrum of 1ES 1553+113, together with NI: Fig. 1 and ED Fig. 1) fills both the disk and an extended thick-disk or halo of our Galaxy.

Thus, if our System-1 and -2 were the analogous of our local OVII absorbers, and this highly ionized gas were confined in the thick-disk or halo of small intervening galaxies, it should also most likely co-exist with much cooler gas and imprint both OVII and strong OI and OII absorption on the spectrum of 1ES 1553+113, which instead is not seen.

**Data Availability.** The entire RGS and COS data used in this work are available in the public XMM-*Newton* and HST archives, namely the "XMM-*Newton* Science Archive (XAS: http://nxsa.esac.esa.int/nxsa-web/#home) and the Milkusky Archive for Space Telescopes (MAST: https://archive.stsci.edu/hst/). In particular, we used XMM-*Newton* ObsIds: 0094380801, 0656990101, 0727780101, 0727780201, 0727780301, 0761100101, 0761100201, 0761100301, 0761100401, 0761100701, 0761110101, 0790380501, 0790380601, 0790380801, 0790380901, 0790381001, 0790381401, 0790381501. We also used the HST-COS G130 and G160 datasets LB4R02010, LB4R02020, LB4R02030, LB4R02040, LB4R02050, LB4R02060, LB4R02070, LB4R02080, LBG803070, LBG803080, LBG803090, LBG8030A0, LBG8030B0, LBG8030C0, LBG8030D0, LBG8030E0, LBG8030F0, LBG8030G0, LBG8030H0, all publicly available in MAST.

**Availability of computer code and algorithm.** The models, hybrid-ionization spectral code and Monte Carlo algorithms used in these work will be made available upon request from the corresponding author FN.

**Extended Data Fig. 1. XMM-*Newton* RGS Spectra of 1ES 1553+113**. Broad-band, unfolded RGS1 (black points and 1σ errorbars) and RGS2 (red points and 1σ errorbars) spectra (in bins with signal to noise per bin ≥20) and best-fitting models (black and red histograms) of the blazar 1ES 1553+113, in physical units. Galactic absorption lines are marked and labeled in blue, while absorption lines from our WHIM System-1 and -2 are marked and labeled in green and magenta, respectively.

**Extended Data Fig. 2. XMM-*Newton* RGS Spectra of 1ES 1553+113**. Normalized raw RGS1 (black points and 1σ errorbars) and RGS2 (red points and 1σ errorbars) data (in bins with signal to noise per bin ≥30) of the blazar 1ES 1553+113, in the wavelength interval λ=26–32 Å. Thick dashed curves are RGS1 (black) and RGS2 (red) best-fitting model folded through the RGSs' response functions. Thin solid curves at the bottom of the graph are RGS1 (black) and RGS2 (red) effective areas (in arbitrary units), showing instrumental features due to cool-pixels in the dispersing detectors. Of the 5 absorption lines shown, only the weak OVII He-β at $z_X$=0.4339, and only in the RGS1, can be affected by the presence of an instrumental feature.

**Extended Data Fig. 3. Assessing Systematics in the RGS Spectrum of 1ES 1553+113.** Outcome of a Monte Carlo procedure consisting in evaluating 10,000 times the single-line statistical significance of an unresolved (i.e. broadened to the instrument Line Spread Function) Gaussian added to the best-fitting continuum-plus-line model of the RGS spectrum of 1ES 1553+113 (see Methods). At each run, the line position is frozen at a random position between 8-33 Å, and its EW is allowed to vary freely from negative (emission) to positive (absorption). The data distribution is symmetric but slightly flatter than the red curve showing the expected Normal distribution for a standard deviation of unity. The Normal distribution that best fits our Monte Carlo results, has a standard deviation of 1.15 (green curve).

**Extended Data Fig. 4. Constancy of the z=0.4339 OVII Absorber.** 2015 (black) and 2017 (red) 800 ks and 950 ks, respectively, RGS spectra of 1ES 1553+113 centered around the OVII He-α (Fig 4a) and He-β (Fig 4b) transitions of System-1. The two lines are consistent with no variability between the two epochs, within their 1σ errors (ED Table 2).

**Extended Data Fig. 5. Galaxy photometric redshifts.** Histogram of photometric redshifts of *r'*>23.5 galaxies within cylindrical volumes with base radii 500 kpc at each redshift interval) and line-of-sight depth Δz=0.075 (the 1σ redshift accuracy) centered on the line of sight to the blazar 1ES 1553+113. The black curve is the average number of galaxies with *r'*>23.5 expected within the explored volumes at each redshift bin, based on the galaxy logN-logS from [11] (where we use a common, but conservative, (B–*r'*) = 1 for all galaxy types).

**Extended Data Fig. 6. *r'*>23.5 galaxies surrounding WHIM System-1.** OSIRIS-GTC mosaic (in band *r'*) of the 12x12 arcmin$^2$ field surrounding the line of sight to 1ES 1553+113 (circled in magenta). Green circles have 500 kpc and 1.75 Mpc radii at z=0.4125. Thin yellow circles highlights the positions of galaxies with photometric redshift estimates within the z=0.375–0.450 interval, while thicker circles highlight spectroscopic redshifts: solid are confirmed and dashed unconfirmed.

**Extended Data Fig. 7. *r'*>23.5 galaxies surrounding WHIM System-2.** Same as ED Fig. 5 but for System-2, in the z=0.300–0.375 interval.

**Extended Data Table 1. Absorption Lines of System-1 and -2.**

**Extended Data Table 2. Lack of Variability of the OVII Absorbers over Two Years**

**Extended Data Table 3. Physics and Chemistry of System-1 and -2.**

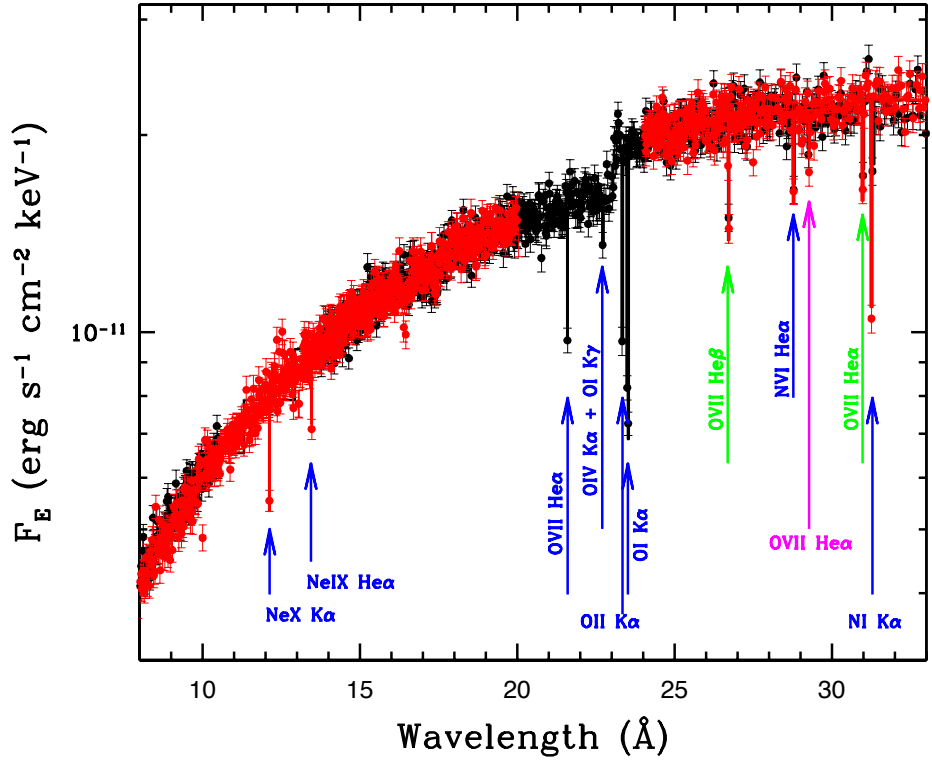

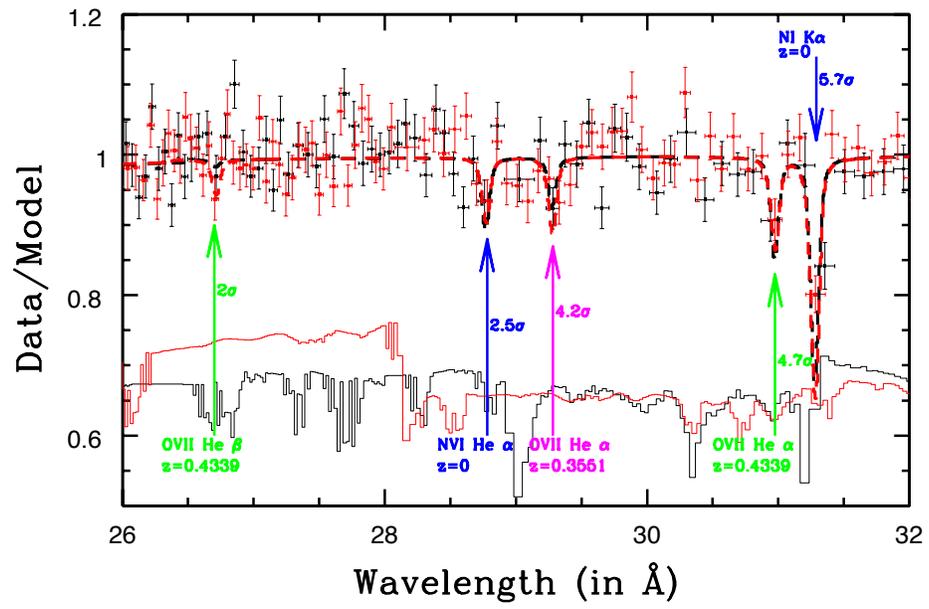

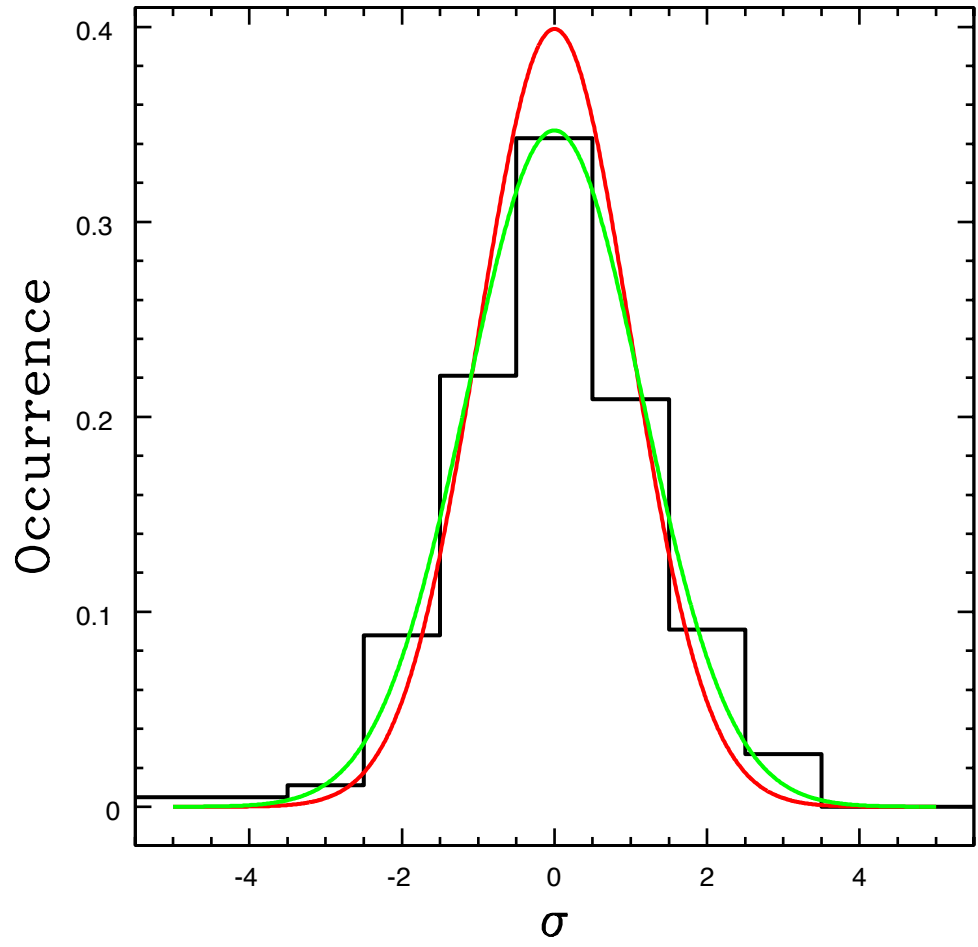

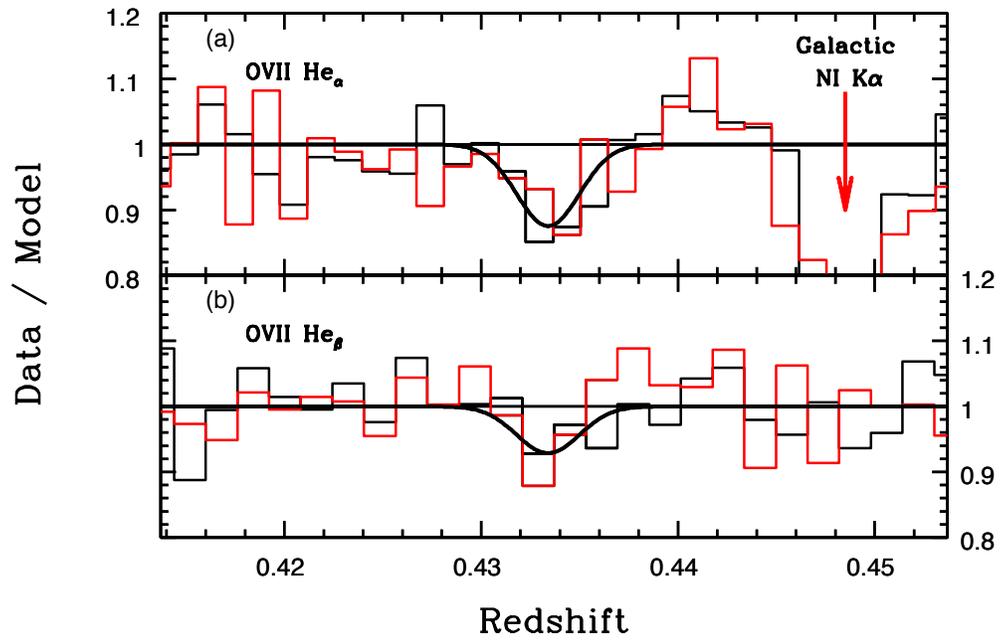

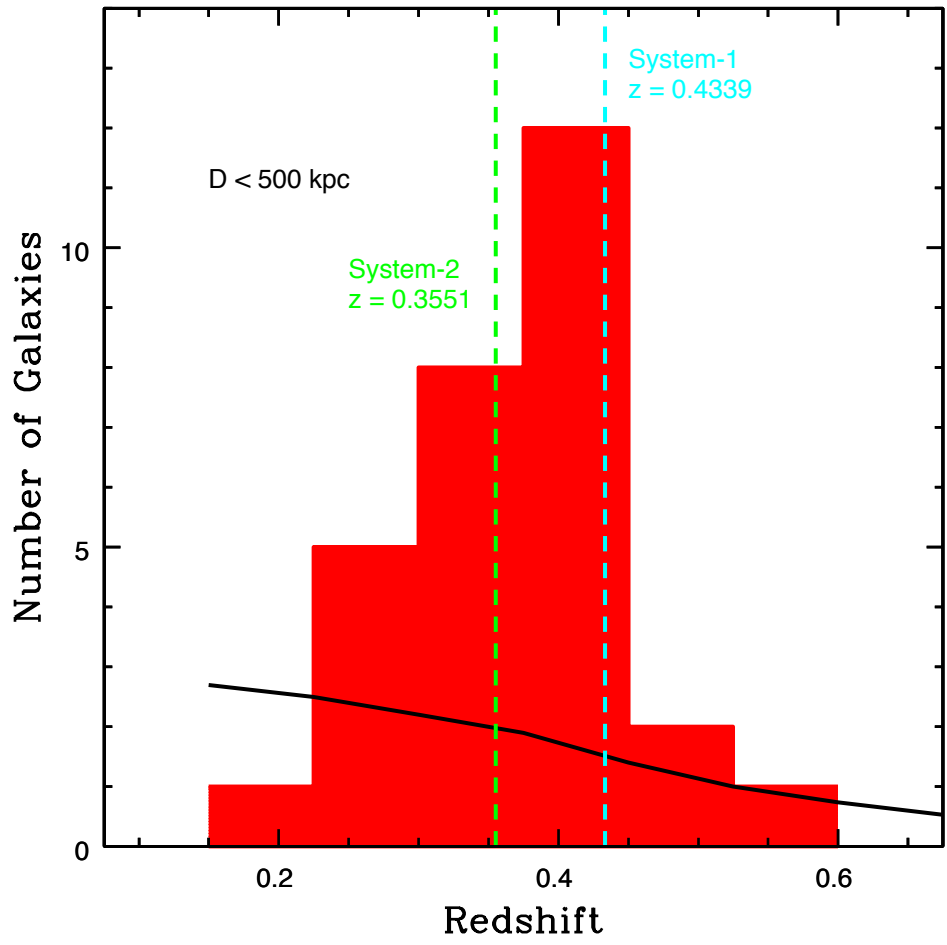

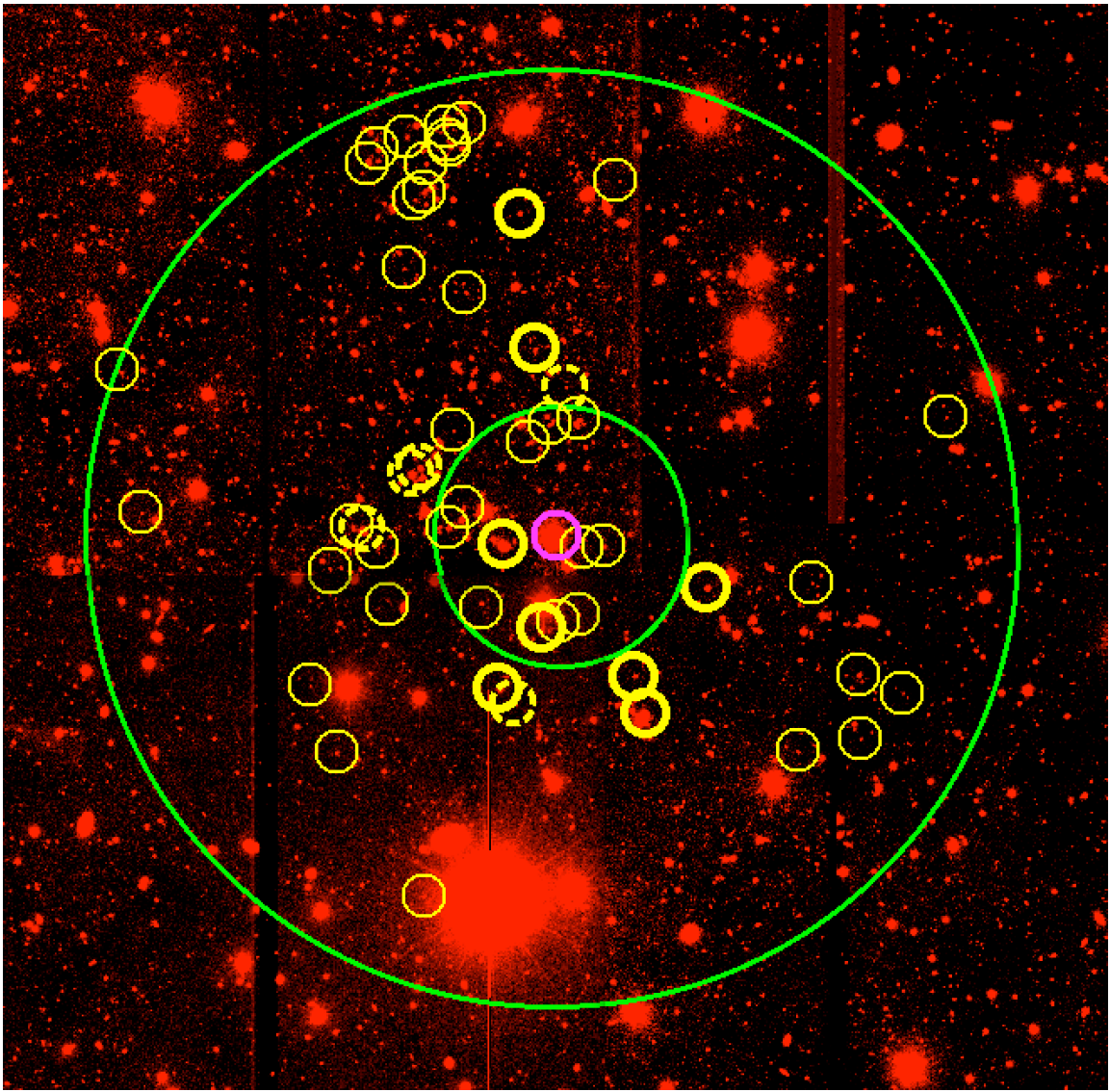

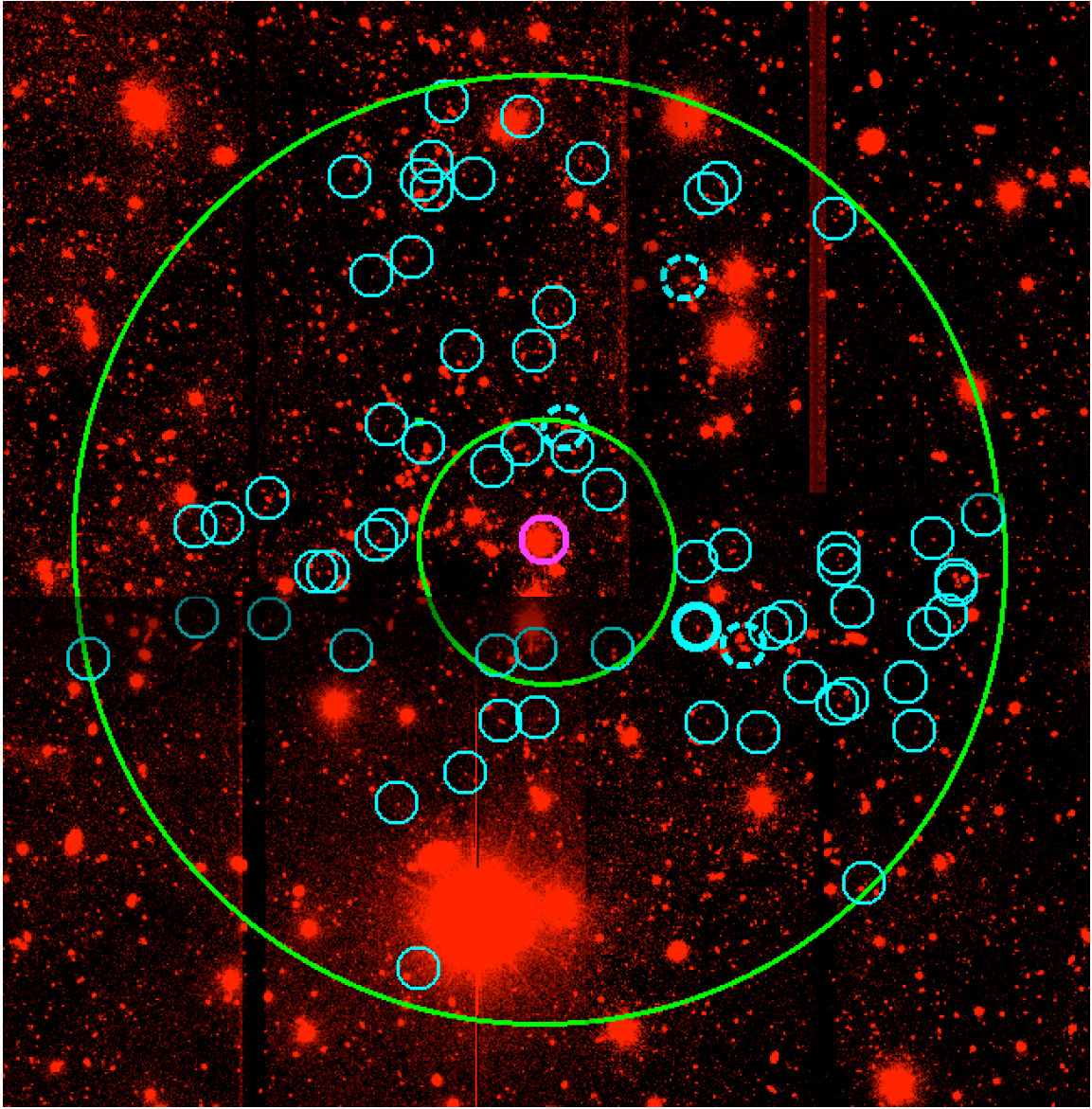

| Wavelength (Å) | Id | Redshift | b (km s$^{-1}$) | EW$_{obs}$ (mÅ) | Significance (σ) |
|---|---|---|---|---|---|
| 30.975±0.017 | OVII He-α | 0.4339±0.0008 | NA | 14.7±3.1 | 4.1–4.7 |
| 26.69±0.09 | OVII He-β | 0.4326±0.0048 | NA | **4.4$^{+2.7}_{-2.2}$** | 1.7–2.0 |
| 1742.18–1744.24 | HI Ly-α | 0.4331–0.4347 | 90–220 | <110 | 3.0 u.l. |
| 1478.86–1480.51 | OVI 2s2p$_{1/2}$ | 0.4331–0.4347 | 55–128 | <30 | 3.0 u.l. |
| 29.27$^{+0.01}_{-0.03}$ | OVII He-α | **0.3551$^{+0.0003}_{-0.0015}$** | NA | **10.5$^{+2.9}_{-2.5}$** | 3.7–4.2 |
| 1645.53–1647.72 | HI Ly-α | 0.3536–0.3554 | 96–190 | <98 | 3.0 u.l. |
| 1396.82–1398.68 | OVI 2s2p$_{1/2}$ | 0.3536–0.3554 | 57–113 | <49 | 3.0 u.l. |

| Epoch | Transition | System | EW$_{obs}$ (in mÅ) |
|---|---|---|---|
| 1 | OVII He-α | 1 | 17±5 |
| 2 | OVII He-α | 1 | 15±5 |
| 1 | OVII He-β | 1 | 7±4 |
| 2 | OVII He-β | 1 | 8±3 |
| 1 | OVII He-α | 2 | 8±5 |
| 2 | OVII He-α | 2 | 12±4 |
| 1 | NI Kα | Gal | 39±7 |
| 2 | NI Kα | Gal | 37±6 |

| System | T (10⁶ K) | N$_O$ (10¹⁵ cm⁻²) | N$_H$(Z/Z$_\odot$)⁻¹ (10¹⁹ cm⁻²) | Z (Z$_\odot$) |
|---|---|---|---|---|
| 1 | 1.2±0.4 | $7.8^{+3.9}_{-2.4}$ | $1.6^{+0.8}_{-0.5}$ | ≥0.1 |
| 2 | 0.95±0.45 | $4.4^{+2.4}_{-2.0}$ | $0.9^{+0.5}_{-0.4}$ | ≥0.1 |